\documentclass[twoside,twocolumn]{article}
\usepackage{amsmath,amssymb}
\usepackage{graphicx}

\begin{document}

\twocolumn[
  \begin{@twocolumnfalse}
\noindent\LARGE{\textbf{A guide to lifting aperiodic structures}}
\vspace{0.6cm}

\noindent\large{\textbf{Michael Baake\textit{$^{1}$}, David \'{E}cija$^{2}$ and
Uwe Grimm\textit{$^{3}$}}}\vspace{0.5cm}

 \noindent \normalsize{The embedding of a given point set with
   non-crystallographic symmetry into higher-dimensional space is
   reviewed, with special emphasis on the Minkowski embedding known
   from number theory. This is a natural choice that does not require
   an a priori construction of a lattice in relation to a given
   symmetry group. Instead, some elementary properties of the point
   set in physical space are used, and explicit methods are described.
   This approach works particularly well for the standard symmetries
   encountered in the practical study of quasicrystalline phases.  We
   also demonstrate this with a recent experimental example, taken
   from a sample with square-triangle tiling structure and
   (approximate) twelvefold symmetry. }
\vspace{1cm}
 \end{@twocolumnfalse}
  ]

\footnotetext[1]{\textit{Fakult\"{a}t f\"{u}r Mathematik, 
       Universit\"{a}t Bielefeld,
       Postf.~100131, 33501 Bielefeld, Germany. 
       E-mail: mbaake@math.uni-bielefeld.de}}
\footnotetext[2]{\textit{IMDEA Nanociencia,  
  C/Faraday, 9, Campus Universitario de Cantoblanco,
  28049 Madrid, Spain.
  E-mail: david.ecija@imdea.org}}
\footnotetext[3]{\textit{School of Mathematics and Statistics, 
       The Open University,
       Walton Hall, Milton Keynes MK7 6AA, United Kingdom. 
       Email: uwe.grimm@open.ac.uk}}

\section{Introduction}

Ever since the discovery of quasicrystals \cite{Dan}, one standard
approach to the investigation of direct images of aperiodic structures
consists of `lifting' a set of positions in two- or three-dimensional
space into a higher-dimensional space, sometimes referred to as
`superspace'. For instance, this could be a set of positions obtained
from electron miscroscopy of a thin slice of a quasicrystal, or from
an STM image of a surface. The lift then provides important
information about the structure of the quasicrystal in terms of a cut
and project description.  From a mathematical point of view, such data
sets are represented as point sets in space (for instance as sets of
atomic or cluster positions), and we are interested in lifting such a
point set into a higher-dimensional space in a suitable way that
reveals the underlying structure.

This approach is particularly useful if the lifted positions come to
lie on a lattice in the higher-dimensional space, so the lift produces
an embedding of the point set into a lattice. Here, a \emph{lattice}
$\varGamma$ in $d$-dimensional real space $\mathbb{R}^{d}$ is defined
as the integer span of a set of $d$ linearly independent vectors
$e_{i}$, $1\le 1\le d$, so that
\[
\begin{split}
  \varGamma & = \biggl\{\sum_{i=1}^{d}n_{i}e_{i}
      \;\big|\;  \text{all } n_{i}\in\mathbb{Z}\biggr\}
  \;\; \text{and} \\
  \mathbb{R}^{d} & =\biggl\{\sum_{i=1}^{d}x_{i}e_{i}
      \;\big|\; \text{all } x_{i}\in\mathbb{R}\biggr\} .
\end{split}
\]
In the literature, one will often find such lifts described for
situations where the higher-dimensional lattice is known in advance.
This may give the impression that, in order to apply this approach to
an observed set of positions, one has to choose a lattice at the
start. However, it may not be obvious what lattice to choose, which
asks for some canonical choice.

In fact, it is possible to employ an \emph{intrinsic} approach where
the lattice is constructed from the observed set of positions
directly, for instance by recovering the `missing part' of the
higher-dimensional coordinates. That this is indeed possible follows
from a non-trivial theorem in \cite{BM04}. It is the goal of this
paper to demonstrate that the underlying construction, in many relevant
situations, is feasible and actually surprisingly simple.

This approach is not new -- in fact, it is pretty much the standard
way it is done in parts of mathematics, in particular in number
theory.  Since the quasicrystal structures that are observed do have a
close connection with these number-theoretic structures, it seems
worth-while to explain the connection and the resulting methods in the
context of quasicrystalline point sets. In this sense, the present
article can be seen as a pedagogic attempt to simplify the handling of
point sets with certain practically relevant symmetries
\cite{Steu04}. As such, it is a continuation of core material from the
recent monograph \cite{TAO}, which also serves as our main reference
for further examples and various mathematical details.

Below, we describe how the lifting can be done, both in theory and in
practice. We start with simple model systems in one dimension, and
then discuss planar examples with eight- and twelvefold
symmetry. Finally, we apply our method to an experimental dataset.

\section{One-dimensional examples}\label{sec:1d}

Let us start with a classic example which is based on the symbolic
\emph{substitution} rule $a\mapsto aba$, $b\mapsto a$ for the binary
alphabet $\{a,b\}$.  Considering $a$ and $b$ as two intervals of
lengths $\lambda=1+\sqrt{2}$ and $1$, the corresponding geometric
\emph{inflation} rule is\smallskip
\[
   \includegraphics[width=\columnwidth]{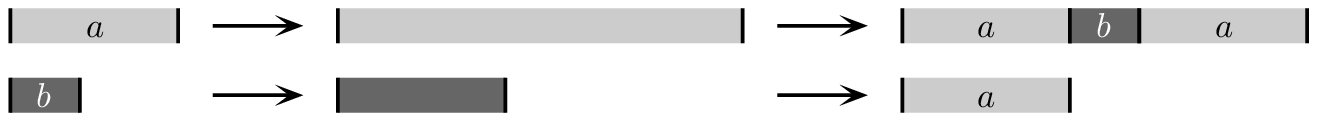}
\]
which maps an interval of type $a$ to three consecutive intervals
$aba$, and the interval $b$ to $a$; see \cite[Def.~4.8]{TAO} for a
more detailed description of substitution versus inflation rules. We
start from a pair of intervals of type $a$ (which, on the symbolic
level, is a legal word of length two because it appears in the second
iterate of the letter $a$) with the origin as its common vertex
point. A repeated application of the inflation rule produces a
one-dimensional tiling of the real axis by intervals of type $a$ and
$b$, according to the symbolic sequence
\[
\begin{split}
   &a|a \,\mapsto\, aba|aba \,\mapsto\, abaaaba|abaaaba \,\mapsto \\
      &abaaabaabaabaaaba|abaaabaabaabaaaba \,\mapsto\, \dots 
   \,\longrightarrow\, w
\end{split}
\]
which converges to a sequence $w$ that is fixed by the substitution
rule (and the corresponding tiling is fixed by the inflation rule).
Here, the vertical line denotes the position of the origin, and the
sequence (as well as the corresponding geometric tiling) is clearly
symmetric under reflection in the origin.  The sequence $w$ is often
referred to as the \emph{silver-mean sequence} due to the continued
fraction expansion $[2;2,2,2,2,\ldots]$ of $\lambda=1+\sqrt{2}$.

Now, collating the left endpoints of each interval of type $a$ and the
left enpoints of intervals of type $b$ produces two point sets
$\varLambda_{a}$ and $\varLambda_{b}$. Their union
$\varLambda=\varLambda_{a}\cup\varLambda_{b}\subset\mathbb{R}$ is
called the \emph{silver mean point set}. By construction, it contains
the origin. Because the two interval lengths are $\lambda$ and $1$,
the distance between neighbouring points is either $\lambda$ or $1$,
and all points must be positioned at integer linear combinations of
these two numbers. Hence, $\varLambda\subset L$ where
\[
   L \, = \, \mathbb{Z}[\sqrt{2}\, ]
   \, = \, \{m+n\sqrt{2}\mid m,n\in\mathbb{Z}\}.
\]
The set $L$ is a dense point set in $\mathbb{R}$, and it is the ring
of integers in the quadratic number field $\mathbb{Q}(\sqrt{2}\, )$,
which is the smallest field extension of the rational numbers that
contains $\sqrt{2}$.  This field has a unique non-trivial automorphism
which is algebraic conjugation, defined by $\sqrt{2}\mapsto
-\sqrt{2}$. For a number $x=m+n\sqrt{2}\in L$, we denote its algebraic
conjugate by $x^{\star}=m-n\sqrt{2}$.

\begin{figure}
\centerline{\includegraphics[width=\columnwidth]{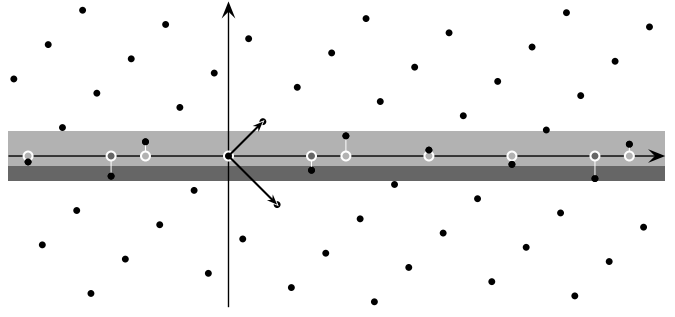}}
\caption{The rectangular lattice $\mathcal{L}$ (black dots) is
  generated by the vectors $(1,1)$ and $\sqrt{2}(1,-1)$. Points in
  $\varLambda_{a}$ (light grey) lift into lattice points within the
  upper (light grey) strip, points in $\varLambda_{b}$ (dark grey)
  lift into lattice points within the lower (dark grey) strip. All
  lattice points within these strips are obtained in this way.}
\label{fig:silverproj}
\end{figure}

Using algebraic conjugation, a natural embedding of $L$ in
$\mathbb{R}^{2}$ is given by
\[
    \mathcal{L} \, = \, \bigl\{(x,x^{\star})\mid 
     x\in L\bigr\}\subset\mathbb{R}^{2},
\]  
which is called the \emph{Minkowski embedding} of $L$; compare
\cite[Ch.~3.4]{TAO}. What does $\mathcal{L}$ look like?  Its elements
(written as row vectors) are of the form
\[
    (m+n\sqrt{2},m-n\sqrt{2}) \, = \,
    m\, (1,1) + n\, (\sqrt{2},-\sqrt{2})
\]
with $m,n\in\mathbb{Z}$, and the representation on the right-hand side
shows that $\mathcal{L}$ is a lattice which is spanned by the two
vectors $(1,1)$ and $\sqrt{2}\, (1,-1)$. These two vectors are clearly
orthogonal, so that $\mathcal{L}$ is a rectangular lattice in the plane;
see Figure~\ref{fig:silverproj} for an illustration.

We thus know that the lift $x\mapsto (x,x^{\star})$ maps points in our
silver mean point set $\varLambda$ into $\mathcal{L}$, so 
\[
    \bigl\{(x,x^{\star})\mid x\in \varLambda\bigr\}
    \, \subset \, \mathcal{L}.
\]
In our example, this turns out to be a very special subset
indeed, as can be seen from Figure~\ref{fig:silverproj}. Explicitly,
all left endpoints of intervals of type $a$ occur at positions $x\in
L$ with $-\frac{2-\sqrt{2}}{2}\le x^{\star}\le \frac{\sqrt{2}}{2}$,
while all left endpoints of intervals of type $b$ are located at
positions $x\in L$ with $-\frac{\sqrt{2}}{2}\le x^{\star}\le
-\frac{2-\sqrt{2}}{2}$.  Note that there are no points in $L$ for
which $x^{\star}$ falls onto the boundaries of these intervals
(because the boundary points are not in $L$ due to the factor
$\frac{1}{2}$), so there is no ambiguity here. The silver mean point
set can now be characterised as the \emph{model set}
(or cut-and-project set)
\[
   \varLambda \, = \, \{x\in L \mid x^{\star}\in W\}
\]  
with the window $W$ being the interval
$W=[-\frac{\sqrt{2}}{2},\frac{\sqrt{2}}{2}]$ which swipes out the
grey strips in Figure~\ref{fig:silverproj}. For the proof of
these statements, we refer to \cite[Ch.~7.1]{TAO}.

At this point, we can lift any subset $S\subset L$, which is a
set in \emph{direct} (or physical) space into \emph{internal}
space as
\[
    S^{\star} \, = \, \{ x^{\star} \mid x \in S \} .
\]
If we do this for the set $\varLambda$, we find that $\varLambda^{\star}$
is a dense subset of the window. Moreover, if we do this for the
finite subsets of the form $\varLambda_{r} := \varLambda \cap [-r,r]$,
the sets $\varLambda^{\star}_{r}$ are finite point sets inside the
window $W$ that fill it out more and more with increasing $r$. In fact,
this is done in a uniform way, which is an important feature known as
\emph{uniform distribution}; see \cite[Prop.~7.3]{TAO} for more. 

Let us note two important things at this point. First, the lift is
done by extracting the missing internal space part $x^{\star}$ from
the known coordinate $x$ in direct space, so that the lift is given by
$x \mapsto (x,x^{\star})$. Second, the lattice and the required
algebraic information is entirely obtained from the set $\varLambda$,
respectively from the set $L$ that emerged from $\varLambda$ via
integer linear combinations. The correct identification of $L$ from
$\varLambda$ can be a little more delicate than in our example at
hand, as is well-known from examples such as the Penrose tiling
vertices; compare \cite[Ex.~7.11]{TAO}. Since we will not meet such
cases below, we suppress the necessary identification of the
\emph{limit translation module} and refer the reader to
\cite[Sec.~5.1.2]{TAO} for further details.
 
Note that, in the construction above, we did not start from a given
lattice, but constructed $\mathcal{L}$ as the Minkowski embedding of
the underlying arithmetic structure of the point set. In this sense,
this is a natural embedding, but we are still free to modify the
choice of lattice by changing the relative scale between the direct
and the internal space. Concretely, one could also use
\[
    \mathcal{L}_{\alpha} \, := \, 
    \bigl\{ (x, \alpha x^{\star}) \mid x \in L \bigr\} 
\]
for any positive $\alpha$, so that our previous choice satisfies
$\mathcal{L} = \mathcal{L}^{}_{1}$. In principle, even negative
$\alpha$ can be used, but since this only results in a reflection in
the $x$-axis, we restrict our attention to $\alpha > 0$. The spanning
vectors of $\mathcal{L}_{\alpha}$ can now be chosen as $b_{1} =
(1,\alpha)$ and $b_{2} = (\sqrt{2}, -\alpha \sqrt{2} )$. For special
choices of $\alpha$, the lattice $\mathcal{L}_{\alpha}$ will actually
be a (scaled) square lattice. For instance, this happens for $\alpha =
\lambda$ (where $\mathcal{L}^{}_{\lambda}$ is then spanned by
$b^{}_{1} = (1,\lambda)$ and $b^{\prime}_{2} = b^{}_{1} + b^{}_{2} =
(\lambda,-1)$), but also for $\alpha = {1/\lambda} = \lambda -
2$ (where $\mathcal{L}^{}_{1/\lambda}$ is spanned by $b^{}_{1} =
(1,\lambda-2)$ and $b^{\prime}_{2} = b^{}_{2} - b^{}_{1} =
(\lambda-2,-1)$).

Although any of these choices seems `nice' in the sense that the
square lattice has a higher symmetry than the original Minkowski
embedding, it should be emphasised that this symmetry is perhaps
appealing, but of no relevance to the problem at hand. This is so
because the relative scale between direct and internal space is 
just a number, without \emph{any} physical meaning.

A completely analogous situation emerges for the well-known
Fibonacci inflation rule\smallskip
\[
    \includegraphics[width=0.75\columnwidth]{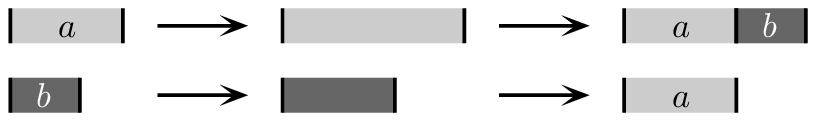}
\]
with intervals of length $\tau$ and $1$ as prototiles. Here, the
dense point set spanned by the positions of the left endpoints
is $L = \mathbb{Z} [\tau]$, the ring of integers in the quadratic
field $\mathbb{Q} (\sqrt{5})$, and the non-trivial field automorphism is
defined by $\sqrt{5} \mapsto -\sqrt{5}$, which means $\tau \mapsto
\tau^{\star} = \frac{-1}{\tau} = 1 - \tau$ and hence 
\[
    m + n \tau \; \longmapsto \;
    (m + n \tau )^{\star} = \, m+n - n \tau  .
\]

Here, the Minkowski embedding is spanned by $(1,1)$ and $(\tau,
1-\tau)$, which is not even a rectangular lattice; see
\cite[Fig.~3.3]{TAO} for an illustration. As before, by scaling
internal space relative to direct space, one can turn the embedding
lattice into a square lattice, as discussed in
\cite[Rem.~3.4]{TAO}. Still, the same comment from above applies,
meaning that such a rescaling of internal space bears no physical
relevance.

\section{Eightfold symmetric tilings}

\begin{figure}
\centerline{\includegraphics[width=0.9\columnwidth]{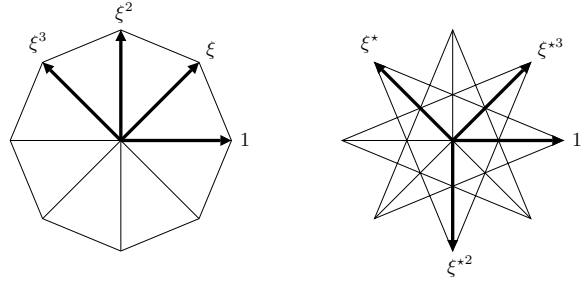}}
\caption{The regular $8$-star in direct space (left panel) with
  $\xi=\xi_{8}=\mathrm{e}^{2\pi\mathrm{i}/8}$ and its $\star$-image in
  internal space (right panel). Here, $\xi^{\star}=\xi^{3}$ and 
$1^{\star}=1$, as well as $(\xi^{n})^{\star}_{}=(\xi^{\star})^{n}_{}$.}
\label{fig:abstars}
\end{figure}

The basic object for eightfold symmetry is the regular $8$-star, as
shown in the left panel of Figure~\ref{fig:abstars}, where it is
natural to use vectors of length $1$. Identifying $\mathbb{R}^{2}$
with $\mathbb{C}$ as usual, the $8$-star is nothing but the star of
all $8$th roots of unity, that is the eight solutions of the equation
$z^{8}-1=0$. Let $\xi^{}_{8}$ be a \emph{primitive} solution (meaning
$\xi^{n}_{8}=1$ only holds for integers that are divisible by $8$),
for instance $\xi^{}_{8} = \mathrm{e}^{2 \pi \mathrm{i}/8}$ to be
explicit (the other primitive solutions being $\xi^{3}_{8}$,
$\xi^{5}_{8}$ and $\xi^{7}_{8}$).

The analogue of the dense point set $L$ from the previous section is
$L^{}_{8}=\mathbb{Z} [\xi^{}_{8} ]$, the ring of integers in the
cyclotomic field $\mathbb{Q} [\xi^{}_{8} ]$; see
\cite[Sec.~2.5.2]{TAO} for an introduction in our context. Any
element of $L^{}_{8}$ is an integer linear combination of
$1,\xi^{}_{8},\xi^{2}_{8}, \dots , \xi^{7}_{8}$, but it turns out that
the first four of them suffice, so
\[
    L^{}_{8} \, = \, \{ m^{}_{0} 1 + m^{}_{1} \xi^{}_{8}
    + m^{}_{2} \xi^{2}_{8} + m^{}_{3} \xi^{3}_{8} \mid
    \text{all } m_{i} \in \mathbb{Z} \} .
\]
Alternatively, one also has that
\[
     L^{}_{8} \, = \, \bigl\{ \alpha^{}_{0} 1 + \alpha^{}_{1} \xi^{}_{8} \mid
     \text{all } \alpha_{i} \in \mathbb{Z} [\sqrt{2}\, ] \bigr\} ,
\]
which means that $L^{}_{8}$ is a $\mathbb{Z}$-module of rank $4$ and,
at the same time, a $\mathbb{Z} [\sqrt{2}\, ]$-module of rank $2$. The
latter property lines up with $L^{}_{8}$ being a dense subset of
$\mathbb{C} \simeq \mathbb{R}^{2}$, while the former tells us that a
lift to a lattice will need $\mathbb{R}^{4}$. Let us thus turn to the
construction of the lattice.

\begin{figure}
\centerline{\includegraphics[width=\columnwidth]{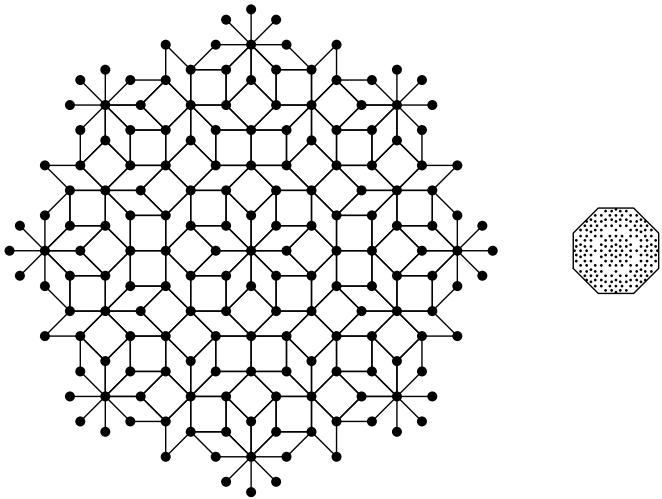}}
\caption{An $8$-fold symmetric patch of the Ammann--Beenker tiling
  (left panel) and the lift of its vertex point set to internal space
  via the $\star$-map for the $\mathbb{Z}$-module $L^{}_{8}$ (right
  panel); see text for details.}
\label{fig:ab}
\end{figure}

The crucial point is the selection of the $\star$-map. In our context,
it has to be one of the field automorphisms of
$\mathbb{Q}(\xi^{}_{8})$; see \cite[Secs.~2.5.2 and 3.4.2]{TAO} for
details. Clearly, it can neither be complex conjugation nor the
trivial one, which leaves us with the choices $\xi^{}_{8}\mapsto
\xi^{3}_{8}$ or $\xi^{}_{8}\mapsto \xi^{5}_{8}$, together with the
unique extension to a field automorphism. Either choice is fine. Let
us use the first one for convenience, so that
$(\xi_{8}^{n})^{\star}_{}=(\xi_{8}^{\star})^{n}_{}=\xi_{8}^{3n}$ for
$n\in\mathbb{Z}$. The effect on the regular $8$-star is shown in the
right panel of Figure~\ref{fig:abstars}.  Now, the Minkowski embedding
gives the lattice
\[
    \mathcal{L}^{}_{8}\, = \, 
    \bigl\{(x,x^{\star})\mid x\in L^{}_{8}\bigr\} \, 
    \subset \, \mathbb{R}^{4}.
\]
One can show that $\mathcal{L}^{}_{8}$ is a scaled (by a factor of
$\sqrt{2}\,$) and rotated version of the integer lattice
$\mathbb{Z}^{4}$; see \cite[Ex.~3.6]{TAO}.

As an example, let us consider the central patch of the $8$-fold
symmetric Ammann--Beenker tiling shown in the left panel of
Figure~\ref{fig:ab}. We assume this patch to be generated by the
inflation rule of \cite[Sec.~6.1]{TAO}, applied to prototiles of unit
edge length. If we give the central vertex the coordinate $0\in
L^{}_{8}$, any other vertex of the patch is an element of $L^{}_{8}$
as well, because every edge has unit length and corresponds to one of
the directions of the regular $8$-star.  Therefore, each vertex can
be indexed by a $4$-tuple of integers,
$(m^{}_{0},m^{}_{1},m^{}_{2},m^{}_{3})$, which represents the point
\[
    x \, = \,  m^{}_{0} 1 + m^{}_{1} \xi^{}_{8} + 
    m^{}_{2} \xi^{2}_{8} + m^{}_{3}\xi^{3}_{8} 
\]
in direct space. Concretely, one finds these $4$-tuples by starting
from the origin and going along the tile edges to the other vertex
points, where each edge with its direction corresponds positively or
negatively to one of the first four vectors of the regular $8$-star;
compare the left panel of Figure~\ref{fig:abstars}. Since the sum of
all vectors of the $8$-star vanishes, the result does not depend on
the path that was chosen.

\begin{figure}
\centerline{\includegraphics[width=\columnwidth]{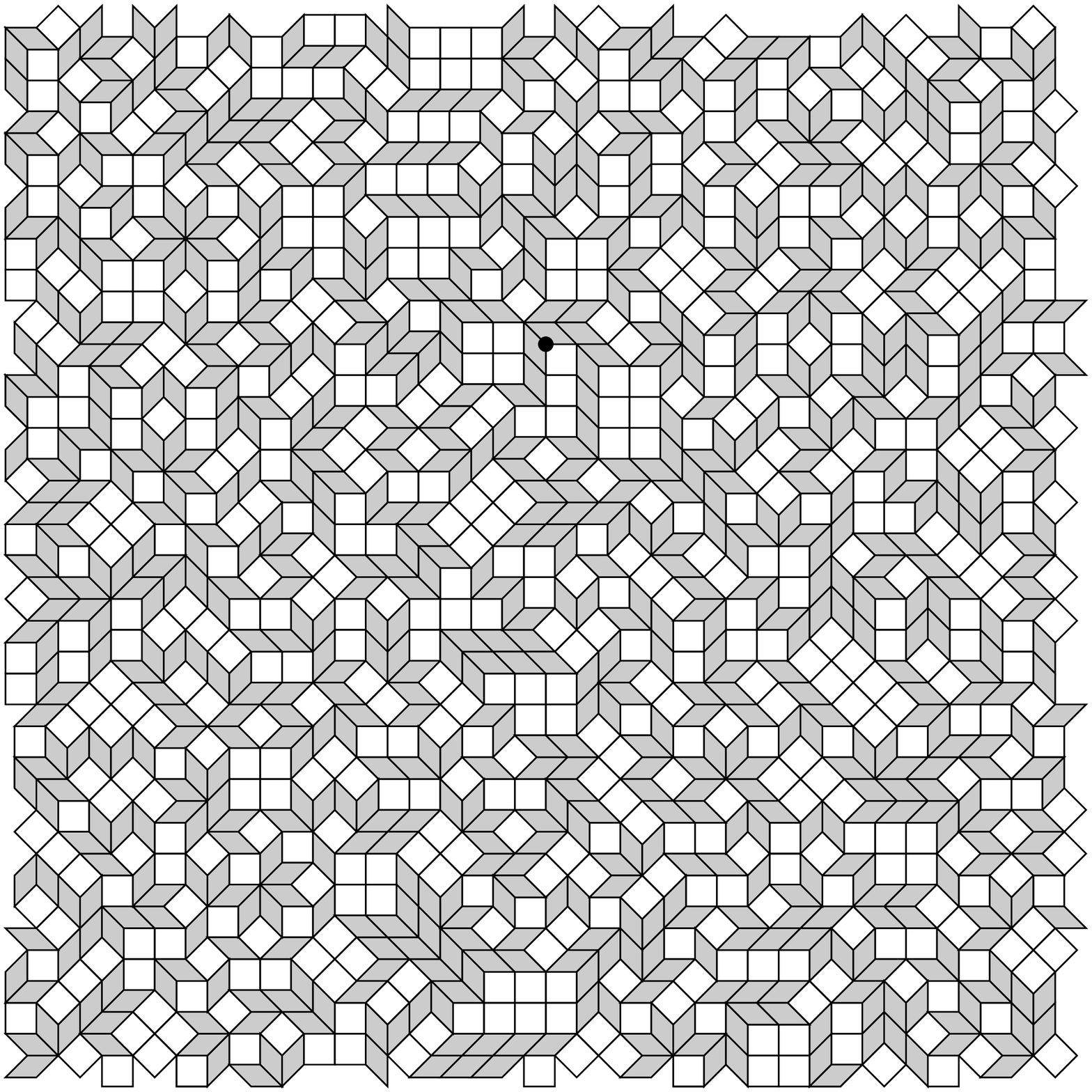}}
\caption{A typical patch of an octagonal random tiling with statistical
$8$-fold symmetry. The marked vertex point has been used as the origin.}
\label{fig:abrand}
\end{figure}

Given any point $(m^{}_{0},m^{}_{1},m^{}_{2},m^{}_{3})$ in $L^{}_{8}$, its
$\star$-image is the point
\[
    x^{\star} \, = \,  m^{}_{0} 1 + m^{}_{1} \xi^{3}_{8} - 
    m^{}_{2} \xi^{2}_{8} + m^{}_{3}\xi^{}_{8} 
\]
in internal space. In terms of the original basis, the $\star$-map
amounts to the mapping
\[
    (m^{}_{0},m^{}_{1},m^{}_{2},m^{}_{3}) \, \mapsto\,
    (m^{}_{0},m^{}_{3},-m^{}_{2},m^{}_{1})
\]
which really is quite simple! Its action on the vertex points of our
Ammann--Beenker patch is shown in the right panel of
Figure~\ref{fig:ab}. The point cloud is the lift of the patch and lies
within a regular octagon of unit edge length, in line with the known
fact that the vertex set of the Ammann--Beenker tiling is a cut and
project set (or model set) for the lattice $\mathcal{L}^{}_{8}$ with
the octagon as its window (or acceptance domain); see
\cite[Sec.~7.3]{TAO} for a detailed exposition. An important feature
of a regular model set such as this one is the fact that the lifted
points, in a natural order according to their distance from the centre
in direct space, are uniformly distributed in the window in internal
space. This is a strong homogeneity property of the system, which
facilitates the calculation of statistical properties as well as a
closed formula for the kinematic diffraction of the point set.

\begin{figure}
\centerline{\includegraphics[width=0.9\columnwidth]{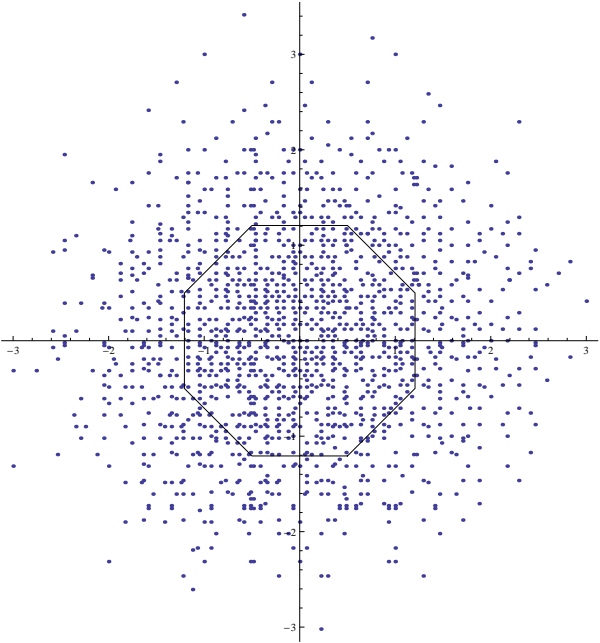}}
\caption{Lift of the vertex points of the octagonal random tiling
  patch of Figure~\ref{fig:abrand} to internal space via the
  $\star$-map. The octagonal window of the perfect Ammann--Beenker
  tiling is shown in the correct relative size.}
\label{fig:ranlift}
\end{figure}

Let us contrast this perfectly ordered structure with its random
tiling counterpart of Figure~\ref{fig:abrand}, which was obtained via
repeated simpleton flips from a perfect patch as described in
\cite[Sec.~11.6.2]{TAO} and references therein. Still, choosing any
vertex point as the origin, all vertex points of the patch are
elements of $L^{}_{8}$, and hence can be lifted via the same
$\star$-map and the method described above. The result is shown in
Figure~\ref{fig:ranlift}, with the chosen origin marked (a different
choice would just result in a shift of the lifted point set). It is
clearly visible that the set of lifted positions extends beyond the
window of the perfect tiling, in agreement with the expectation for
the statistics of random tilings \cite{Henley}.

The very same method, with minor modifications, works for any rhombus
tiling with edges along the directions of a regular $n$-star. In fact,
it also works if we have a set of prototiles with edges of the same
length along such a set of directions. In general, the dimension of
the internal space becomes larger; see \cite[Sec.~7.3]{TAO} for
details.  Here, we restrict our attention to the practically most
important cases where internal space has the same dimension as direct
space. In view of dodecagonal quasicrystals \cite{INF85,CKH98} and
various recent developments, the possibly most relevant example is
that of $12$-fold symmetry, with square-triangle tilings featuring
prominently; see \cite{Zeng,Ron,surf,soft} and references therein.

\section{Square-triangle tilings}

\begin{figure}[h]
\centerline{\includegraphics[width=0.9\columnwidth]{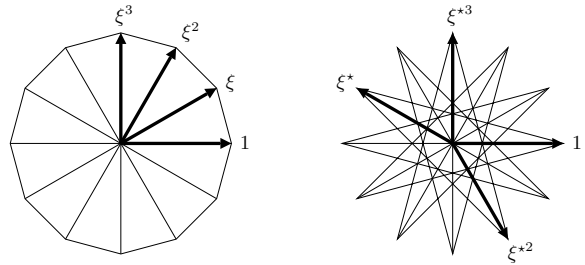}}
\caption{The regular $12$-star in direct space (left panel) with
  $\xi=\xi_{12}=\mathrm{e}^{2\pi\mathrm{i}/12}$ and its $\star$-image
  in internal space (right panel), where $\xi^{\star}=\xi^{5}$. Note
  that $\xi^{3}=(\xi^{\star})^{3}=\mathrm{i}$.}
\label{fig:12stars}
\end{figure}

The analogue of Figure~\ref{fig:abstars} for $12$-fold symmetry is
given by the regular $12$-star of Figure~\ref{fig:12stars} (left
panel) and its $\star$-image (right panel). The integer span of the
regular $12$-star is
\[
    L^{}_{12} \, = \,  \{ m^{}_{0} 1 + m^{}_{1} \xi^{}_{12}
    + m^{}_{2} \xi^{2}_{12} + m^{}_{3} \xi^{3}_{12} \mid
    \text{all } m_{i} \in \mathbb{Z} \} \, ,
\]
which is once again a $\mathbb{Z}$-module of rank $4$. In particular,
one has $\xi^{4}_{12}=\xi^{2}_{12}-1$ and
$\xi^{5}_{12}=\xi^{3}_{12}-\xi^{}_{12}$, while the remaining powers of
$\xi^{}_{12}$ are obtained via multiplication by $-1$ from the powers
so far.

The module $L^{}_{12}$ is also a $\mathbb{Z}[\sqrt{3}\,]$-module of
rank $2$, meaning that
\[
     L^{}_{12} \, = \, \bigl\{ \alpha^{}_{0} 1 + \alpha^{}_{1} \xi^{}_{12} 
     \mid \text{all } \alpha_{i} \in \mathbb{Z} [\sqrt{3}\, ] \bigr\} ,
\]
see \cite[Sec.~2.5.2]{TAO} for details. As in our previous example,
the $\star$-map is one of the suitable field automorphisms of the
cyclotomic field $\mathbb{Q}(\xi^{}_{12})$. Here, one has the choice
between $\xi^{}_{12}\mapsto\xi^{5}_{12}$ and
$\xi^{}_{12}\mapsto\xi^{7}_{12}$, where we have selected the former.
The action on the $12$-star is shown in Figure~\ref{fig:12stars}.

\begin{figure}
\centerline{\includegraphics[width=\columnwidth]{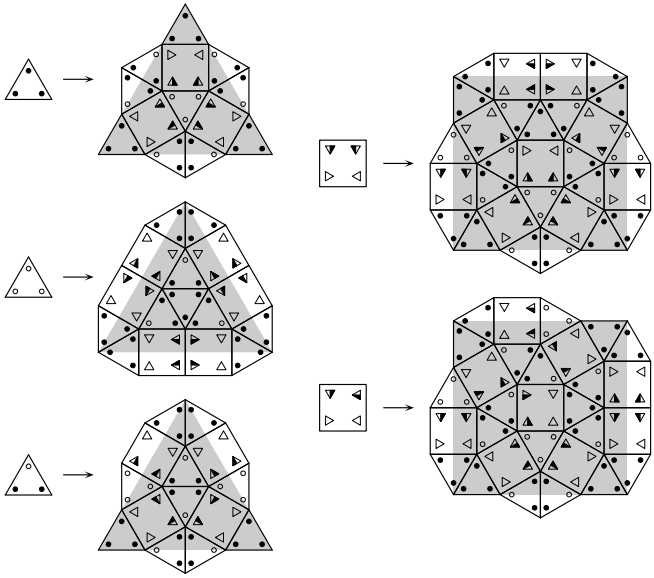}}
\caption{Schlottmann's pseudo inflation rule for a square-triangle tiling
formulated via five decorated prototiles (up to similarity).}
\label{fig:schlotti}
\end{figure}

\begin{figure}
\centerline{\includegraphics[width=\columnwidth]{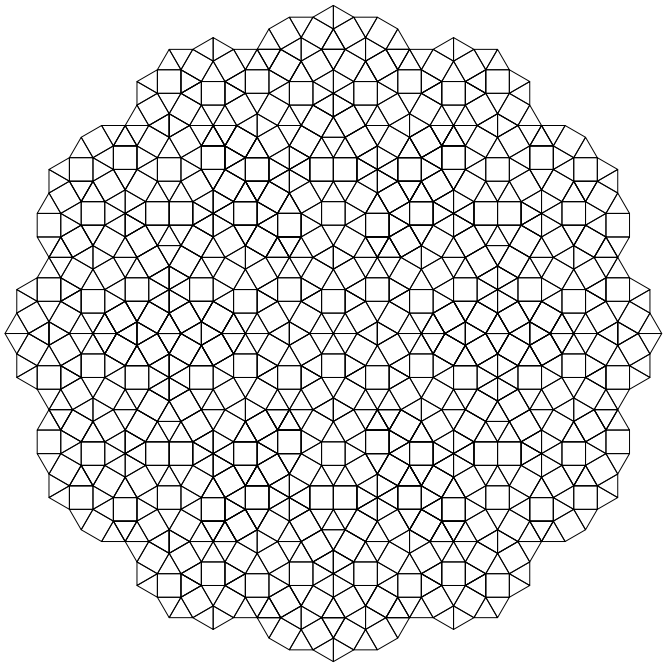}}
\caption{A patch of the square-triangle tiling obtained via the pseudo
  inflation rule of Figure~\ref{fig:schlotti}. Note that the
  decorations are required to construct the patch, but have been
  omitted in this figure.}
\label{fig:12til}
\end{figure}

In this case, the $\star$-image of a 
point $(m^{}_{0},m^{}_{1},m^{}_{2},m^{}_{3})$ in $L^{}_{12}$ is given by
\[
    x^{\star} \, = \,  (m^{}_{0}+m^{}_{2}) 1 - m^{}_{1} \xi^{}_{12} - 
    m^{}_{2} \xi^{2}_{12} + (m^{}_{1}+m^{}_{3})\xi^{3}_{12} 
\]
in internal space. Hence, the $\star$-map acts on the $4$-tuples of
integer coordinates as
\[
    (m^{}_{0},m^{}_{1},m^{}_{2},m^{}_{3}) \, \mapsto\,
    (m^{}_{0}+m^{}_{2},-m^{}_{1},-m^{}_{2},m^{}_{1}+m^{}_{3})\, .
\]
We are now prepared to lift arbitrary subsets of $L^{}_{12}$ to
internal space. Let us note that there is a canonical Minkowski
embedding again, leading to the lattice
$\mathcal{L}^{}_{12}=\bigl\{(x,x^{\star})\mid x\in
L^{}_{12}\bigr\}\subset \mathbb{R}^{4}$.  Of course, one could also
use $\bigl\{(x,\alpha x^{\star})\mid x\in L^{}_{12}\bigr\}$ with
$\alpha>0$, which gives us the freedom to select a `nice' lattice in
$4$-space; see \cite[Ex.~3.6 and Rem.~3.5]{TAO} for details. However,
as explained in Section~\ref{sec:1d} above, the parameter $\alpha$ has
no physical relevance whatsoever and is not needed to describe the
lift or the structure in direct space. Therefore, we prefer to
dispense with it altogether for our discussion.

\begin{figure}
\centerline{\includegraphics[width=\columnwidth]{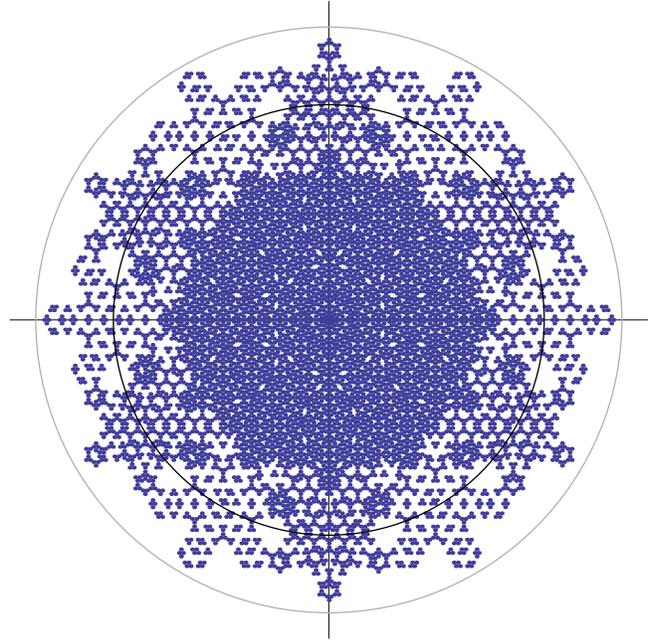}}
\caption{Lift of the vertex points of the next inflation step of the
  square-triangle patch of Figure~\ref{fig:12til} to internal space
  via the $\star$-map. The black circle indicates the size of a
  circular window for a cut and project set of the same density. Note
  that the exact sixfold symmetry is a consequence of the
  corresponding symmetry of the underlying patch, while the $12$-fold
  symmetry of the fractally bounded window will only emerge in the
  infinite size limit. The grey circle indicates the circumcircle of the
  (fractally bounded) window.}
\label{fig:12int}
\end{figure}

This setting can now be applied to $12$-fold rhombus tilings, but also
to examples such as G\"{a}hler's shield tiling, see \cite{Gaehl88} or
\cite[Sec.~6.3.2]{TAO}, or to the large family of square-triangle
tilings.  Let us consider the latter case, and apply the setting to
the $12$-fold symmetric square-triangle tiling of the plane that is
obtained by Schlottmann's pseudo inflation rule of
Figure~\ref{fig:schlotti}; see \cite[Sec.~6.3.1]{TAO} and references
therein for background. A patch of the (undecorated) tiling is shown
in Figure~\ref{fig:12til}. It is known that the vertex points of this
tiling form a cut and project set (or model set), where the window is
a $12$-fold symmetric region in the plane with fractal boundary; see
\cite[Fig.~7.10 and Rem.~7.9]{TAO}. From the inflation rule, one can
calculate that the vertex point set has density
$(3+2\sqrt{3})/6\approx 1.077$. Since the lattice
$\mathcal{L}^{}_{12}$ has density $\frac{1}{3}$ in $4$-space, the area
of the window must be $(3+2\sqrt{3})/2$. This implies that a model set
with a circular window of radius $\sqrt{(3+2\sqrt{3})/2\pi}\approx
1.014$ would be a point set of the same density (which differs in many
positions though).

\begin{figure}
\centerline{\includegraphics[width=\columnwidth]{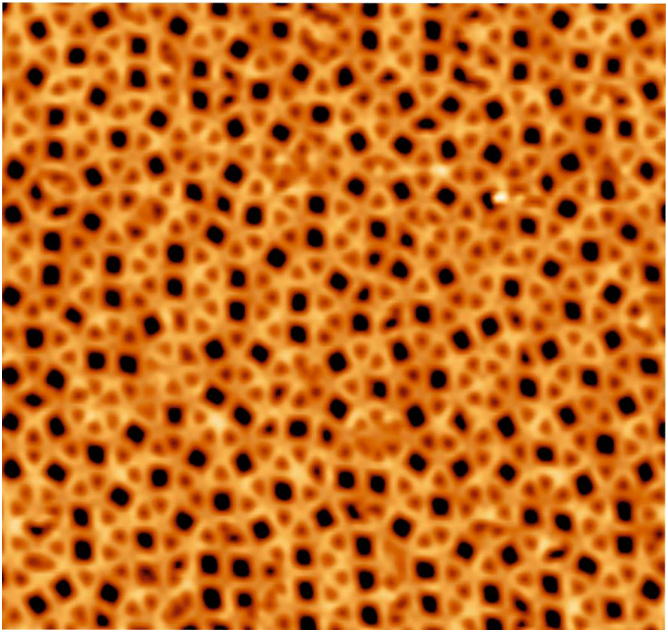}}
\caption{STM image of a metal-organic coordination network \cite{Urgel}.}
\label{fig:expic}
\end{figure}

\begin{figure}
\centerline{\includegraphics[width=\columnwidth]{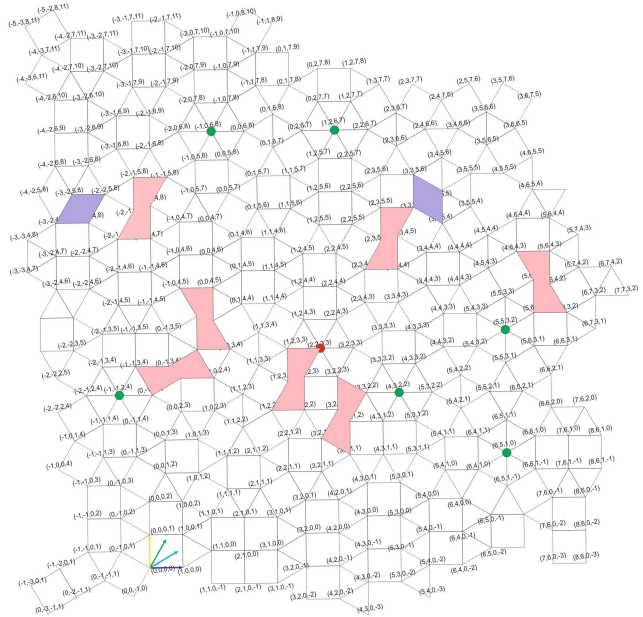}}
\caption{Square-triangle tiling (with defects) obtained by explicit,
slightly idealised coordinatisation from the experimental STM image of
Figure~\ref{fig:expic}. Note that the choice of origin is arbitrary.}
\label{fig:excoord}
\end{figure}

\begin{figure}
\centerline{\includegraphics[width=0.9\columnwidth]{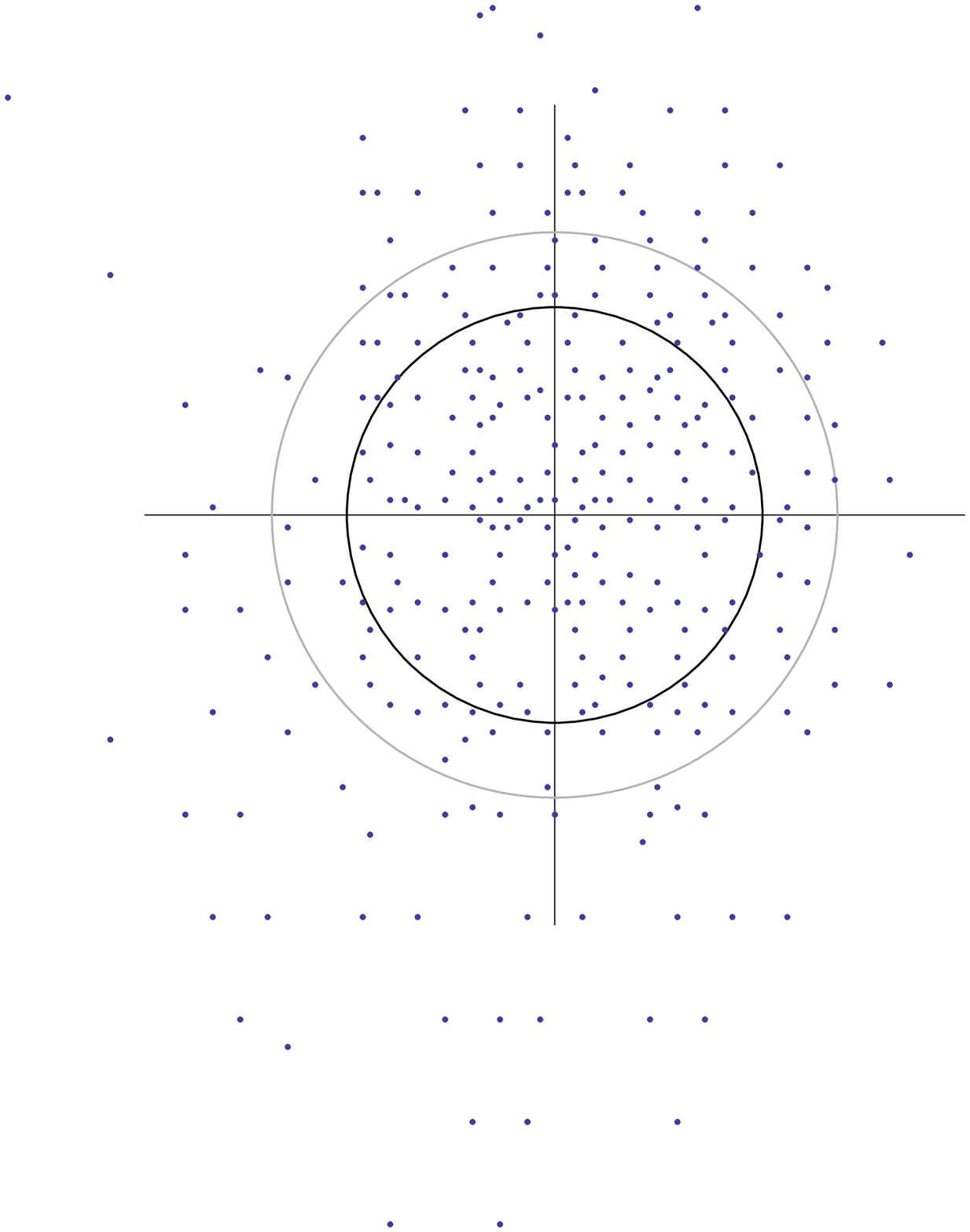}}
\caption{Lift of the vertex points of the experimental tiling of
  Figure~\ref{fig:excoord} to internal space via the $\star$-map. The
  figure is centred in the barycentre of the point cloud.  The two
  circles then exactly correspond to those of Figure~\ref{fig:12int};
  see text for further details.}
\label{fig:expint}
\end{figure}

The result of the lift to internal space via the $\star$-map is shown
in Figure~\ref{fig:12int}. Here, we have started from the patch that
emerges from Figure~\ref{fig:12til} by one additional inflation step,
which has 8623 vertices. For comparison, the circular window mentioned
above is indicated in the figure. While some lifted points fall
outside this circle, they remain within the window of the
square-triangle tiling which is the compact set of
\cite[Fig.~7.10]{TAO}.

\section{Sample application}

Let us finally apply the method to Figure~\ref{fig:expic}, which shows
an experimental STM image \cite{Urgel}. It was obtained by STM
analysis of a metallo-supramolecular network, which is based on
Europium-ligand coordination motifs.  The molecules appear as rod-like
protrusions in the STM data, whereas Eu atoms reside at the
intersection points. We identify distinct coordination nodes which are
interconnected by certain molecular linkers and span an intricate,
fully reticulated metallo-supramolecular network. The individual Eu
centres are surrounded by four, five or six molecules. Moreover, the
Eu vertices and linker backbones are distributed in such a fashion
that the design can be interpreted as a surface tessellation based on
a square-triangle tiling with defects; see Figure~\ref{fig:excoord}
for the result with explicit 4D integer coordinates.

Figure~\ref{fig:expint} shows the corresponding lift to internal space
by our previously described method. The orientations correspond to the
right panel of Figure~\ref{fig:12stars}. The two circles in
Figure~\ref{fig:expint} exactly correspond to those of
Figure~\ref{fig:12int}. Here, the distribution of lifted points
clusters around the barycentre, but is both less regular and more
spread out than the points from a perfect cut and project set. In
particular, one sees a preferred direction. For further
interpretations, we refer to \cite{Urgel}.

\section*{Acknowledgements}

It is our pleasure to thank Johannes Roth for useful hints. We are
grateful to Johannes V.\ Barth, Nian Lin and Jos\'{e} I.\ Urgel for
providing the experimental image from \cite{Urgel}.  This work was
supported by the German Research Council (DFG), within the CRC 701.

\end{document}